\documentclass[debug,overfull,figures,info,cite]{epl}%

\usepackage{latexsym}
\usepackage{amssymb}
\usepackage{graphicx,rotating}

\newcommand{\twofigwidth}{0.48\columnwidth}

\title{Fluctuating Hall resistance defeats the quantized Hall insulator}

\author{Philipp Cain\inst{1} and Rudolf A.\ R\"{o}mer\inst{2}}
\institute{
 \inst{1} Institut f\"{u}r Physik, Technische Universit\"{a}t
Chemnitz, 09107 Chemnitz, Germany\\
 \inst{2} Department of Physics and Centre for Scientific Computing, University of Warwick, Coventry CV4 7AL, United
 Kingdom
}


\begin{document}

\maketitle

\begin{abstract}
Using the Chalker-Coddington network model as a drastically
simplified, but universal model of integer quantum Hall physics,
we investigate the plateau-to-insulator transition at strong
magnetic field by means of a real-space renormalization approach.
Our results suggest that for a fully quantum coherent situation,
the quantized Hall insulator with $R_{\rm H}\approx h/e^2$ is
observed up to $R_{\rm L}\sim 25 h/e^2$ when studying the most
probable value of the distribution function $P(R_{\rm H})$. Upon
further increasing $R_{\rm L}\rightarrow\infty$ the Hall insulator
with diverging Hall resistance $R_{\rm H}\propto R_{\rm
L}^{\kappa}$ is seen. The crossover between these two regimes
depends on the precise nature of the averaging procedure.
\end{abstract}

\section{Introduction}

The integer quantum Hall (QH) transitions are described well in
terms of a series of delocalization-localization transitions of
the electron wavefunction \cite{WeiTP85}.
These universal plateau-plateau transitions are accompanied by a
power-law divergence $\epsilon^{-\nu}$ of the electronic
localization length $\xi$, where $\epsilon$ defines the distance
to the transition for a suitable controlled parameter, e.g.\ the
electron energy \cite{KucMDK00}.
Similarly, it is now conclusively established that plateau-plateau
and insulator-plateau transitions exhibit the same critical
behavior
\cite{GolWSS93,
ShaTSC97,HilSST98,LanPVP02}.

However, the value of the Hall resistance $R_{\rm H}$ in this
insulating phase (at large magnetic field) is still rather
controversial. Various experiments have found that $R_{\rm H}$
remains very close to its quantized value $h/e^2$ even deep in the
insulating regime \cite{ShaTSC97,HilSST98,LanPVP02} with
longitudinal resistance $R_{\rm L} > h/e^2$. This scenario has
been dubbed the {\em quantized Hall insulator}. On the other hand,
theoretical predictions show that a diverging $R_{\rm H}$ should
be expected, i.e., $R_{\rm H} \propto R_{\rm L}^{\alpha}$
\cite{PryA99,ZulS01}. This {\em Hall insulator} is to be expected
at strong disorder or strong magnetic fields.

In fully quantum coherent transport measurements such as in
mesoscopic devices at low temperature, the results clearly show
the paramount influence of quantum interference and the measured
quantities fluctuate strongly \cite{JaiK88}. At magnetic field
$B=0$, the universal conductance fluctuations provide the most
famous example \cite{LeeSF87}. For the QH situation, similarly
reproducible and pronounced fluctuations have been observed
previously
\cite{GalR97,WeyJ98,CobK96,
PelSCS03}, although no complete understanding of their behavior
has yet emerged. Thus for quantum coherent QH samples, the average
values $R_{\rm L}$ and $R_{\rm H}$ become meaningless unless their
full distributions are being taken into account \cite{EntALI95}.
Consequently, different averaging procedures may then yield quite
different estimates for $R_{\rm L}$ and $R_{\rm H}$.

In the present manuscript, we revisit the Hall insulating regime
by using a real-space renormalization group (RG) approach to the
Chalker-Coddington (CC) network model, recently used successfully
to study the energy-level statistics at the QH transition as well
as the influence of long-range correlations in the disorder
potential close to the transition \cite{CaiRSR01}. We will show
when considering the experimentally relevant most probable value
as estimate for $R_{\rm H}$, we find quantized $R_{\rm H}=h/e^2$
for $R_{\rm L} \leq 10 h/e^2$ and increasing values for larger
$R_{\rm L}$ thereby reconciling the experimental results of a
quantized $R_{\rm H}$ available up to $R_{\rm L} =8 h/e^2$
\cite{HilSST98,LanPVP02,PelSCS03} with the theoretical predictions
of the Hall insulator.

\section{Model and RG approach to the QH situation}
\label{sec-RG}

The CC model is based on the microscopic picture of electron
motion in a strong magnetic field and a smooth disorder
potential\cite{ChaC88} when only the semiclassical trajectories of
the guiding center of the cyclotron orbit are important. Assigning
these trajectories to {\em links} and considering saddle points
(SP's) at which different trajectories come closer than the Larmor
radius as {\em nodes} a chiral network can be constructed.
\begin{figure}[t]
  \onefigure[scale=0.5]{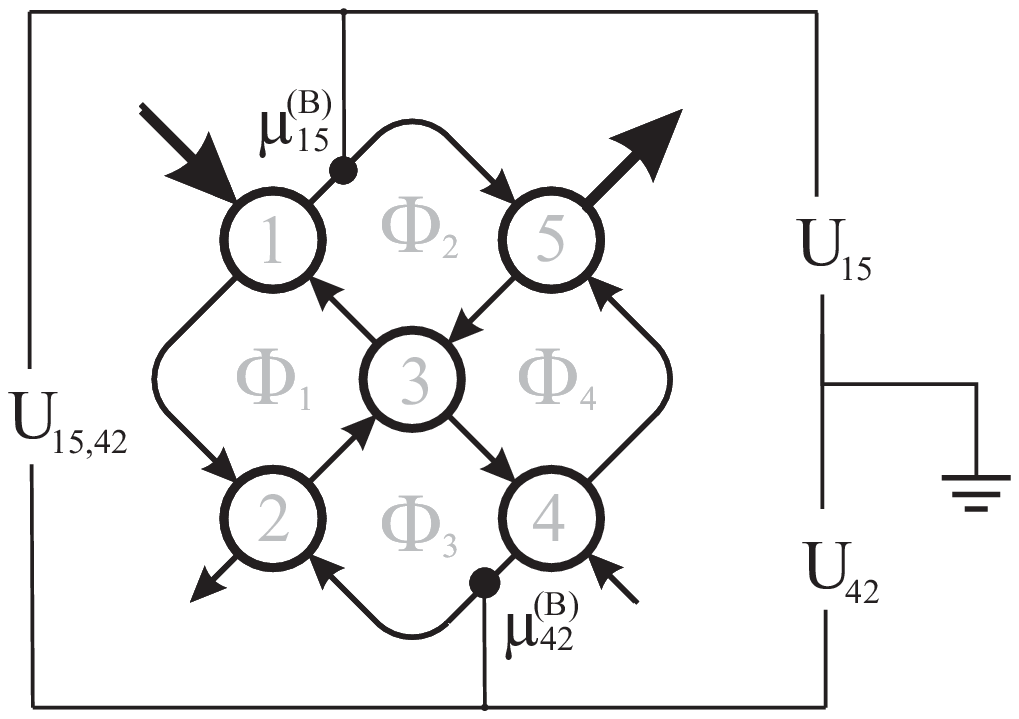}
\caption{\label{fig-RGstruct}
  RG unit of 5 SP's (full circles) and equipotential lines (arrows) on a
  square lattice with Eq.\ (\ref{eq-qhrg}) combining them into a super-SP.
  $\Phi_1, \ldots, \Phi_4$ are the phases acquired by an electron drifting
  along the contours. Arrows show the electron motion for one $B$ direction
  and should be reversed for $-B$.
  The small solid dots indicate the position of the weakly-coupled voltage probes.
  The thin lines denote the different possibilities of measuring
  voltage drops in the present structure for both $B$-field directions.}
\end{figure}

We now apply a real-space RG approach \cite{GalR97,AroJS97} to the
CC network
\cite{WeyJ98,
SinMG00,CaiRSR01,ZulS01}. The RG unit we use is extracted from a
CC network on a regular 2D square lattice as shown in Fig.\
\ref{fig-RGstruct}.  A super-SP consists of five original SP's by
analogy to the RG unit employed
for the 2D bond percolation problem \cite{StaA92
}.
Each SP can be described by two matrix equations relating the
wavefunction amplitudes in incoming and outgoing channels. Between
the SP's an electron travels along equipotential lines, and
accumulates a certain Aharonov-Bohm phase $\Phi$. Different phases
are uncorrelated, which reflects the randomness of the original
potential landscape.  This results in a system of ten linear
equations, the solution of which yields the expression for the
transmission coefficient of the super-SP \cite{GalR97}
\begin{equation}
  \label{eq-qhrg}
  t'= \left | \frac{ t_1 t_5 (r_2 r_3 r_4 e^{\imath\Phi_2} - 1) + t_2
      t_4 e^{\imath (\Phi_3+\Phi_4)} (r_1 r_3 r_5 e^{-\imath\Phi_1} -
      1) + t_3 (t_2 t_5 e^{\imath\Phi_3} + t_1 t_4 e^{\imath\Phi_4}) }
    { (r_3 - r_2 r_4 e^{\imath\Phi_2}) (r_3 - r_1 r_5 e^{\imath
        \Phi_1}) + (t_3 - t_4 t_5 e^{\imath\Phi_4}) (t_3 - t_1 t_2
      e^{\imath\Phi_3}) }\right | \quad .
\end{equation}
Here $t_i$ and $r_i=(1-t_i^2)^{1/2}$ are, respectively, the
transmission and reflection coefficients of the constituting SP's,
$\Phi_j$ are the phases accumulated along the closed loops (see
Fig.\ \ref{fig-RGstruct}).  Equation (\ref{eq-qhrg}) is the RG
transformation, which allows one to generate (after averaging over
$\Phi_j$) the distribution $P(t')$ of the transmission
coefficients of super-SP's using the distribution $P(t)$ of the
transmission coefficients of the original SP's. Since the
transmission coefficients of the original SP's depend on the
electron energy $\varepsilon$, the fact that delocalization occurs
at $\varepsilon = 0$ implies that a certain distribution, $P_{\rm
c}(t)$ --- with $P_{\rm c}(t^2)$ being symmetric with respect to
$t^2=\frac{1}{{2}}$ --- is a fixed point (FP) of the RG
transformation (\ref{eq-qhrg}).  The distribution $P_{\rm c}(G)$
of the dimensionless two-point conductance $G$ can be obtained
from the relation $G=t^2$, so that $P_{\rm c}(G)\equiv P_{\rm
  c}(t)/2t$.
We remark that the classical percolation limit
\cite{StaA92
} is obtained when the classical values
$t \in \{0,1\}$ are used in the RG.

Let us now make contact with the resistances. The {\em
dimensionless} longitudinal resistance $R_{\rm L}$ can be computed
from $G$ via
\begin{equation}\label{eq-RL}
    R_{\rm L} = \frac{|r|^2}{|t|^2} = \frac{1 - |t|^2}{|t|^2} =
    \frac{1}{G} - 1 \equiv R_{\rm 2t} - 1
\end{equation}
with the dimensionless $2$-terminal resistance $R_{\rm 2t}$. In
order to study the Hall resistance for the four-terminal CC model,
we study the difference in (normalized) chemical potentials
$\mu_{15}$ and $\mu_{42}$ at the positions indicated in Fig.\
\ref{fig-RGstruct} \cite{PryA99,ZulS01}. In order to remove any
cross-current amplitudes, the voltage is computed, analogously to
the experimental situation, by subtracting opposite magnetic field
directions. The Hall voltage is then defined as
\begin{equation}\label{eq-VH}
    U_{\rm H} = \frac{1}{2} \left\{ \left[ \mu_{15}(B) - \mu_{42}(B)
    \right] - \left[ \mu_{15}(-B) - \mu_{42}(-B)
    \right] \right\} \quad .
\end{equation}
Consequently, the {\em dimensionless} Hall
resistance is given by
\begin{equation}\label{eq-RH}
    R_{\rm H} = \frac{U_{\rm H}}{G} \quad .
\end{equation}

\section{Numerical RG procedure}
\label{sec-numerics}

In order to find the FP distributions $P_{\rm c}(R_{\rm L})$ and
$P_{\rm c}(R_{\rm H})$ at the QH transition, we start from an
appropriate initial distribution of transmission coefficients,
$P_0(t)$. The RG procedure results in a broad distribution $P_{\rm
c}(G)$ which is peaked at $G\gtrsim 0$ and $G\lesssim 1$
\cite{CaiRSR01}. Any RG flow away from this instable FP towards
the insulating regimes will result in a further increase of one of
these peaks. Therefore, in order to reliably model the insulating
regime, an accurate procedure for reproducing these peaks is
needed. In \cite{CaiRSR01}, we had also shown that a distribution
$Q(z)$ of dimensionless SP heights is related to $P(G)$ via $Q(z)
= P(G) (dG/dz) = \frac{1}{4}\cosh^{-2}(z/2)
P\left[(e^z+1)^{-1}\right]$ and $t=(e^{z}+1)^{-1/2}$. Thus the
peaks in $P(G)$ transform into long but quickly decaying tails in
$Q(z)$ and it is numerically much better to perform the RG for SP
heights $z$ when the physics away from the FP is studied.

We discretize the distribution $Q(z)$ in at least $6000$ bins such
that the bin width is typically $0.01$. Since $z \in ]-\infty,
\infty[$, we have to include lower and upper cut-off SP heights
such that $z \in [z_{\rm low}, z_{\rm up}]$. We use $z_{\rm low}=
-20$, $z_{\rm up}= 40$ for perturbations towards positive $z$ and
vice versa for negative perturbations. This corresponds to
transmission amplitudes $t(z_{\rm low})\approx 10^{-5}$, $t(z_{\rm
up})\approx 1-10^{-9}$ for positive perturbations and vice versa
for negative perturbations. From $Q_0(z)$, we obtain $z_i$,
$i=1,\ldots,5$, compute the associated $t_i$ and substitute them
into the RG transformation (\ref{eq-qhrg}).  The phases $\Phi_j$,
$j= 1,\ldots,4$ are chosen randomly from the interval $\Phi_j \in
[0, 2\pi]$.  In this way we calculate at least $10^{8}$
super-transmission coefficients $t'$ and associated $z'$. At the
next step we repeat the procedure using $Q_1$ as an initial
distribution.  We assume that the iteration process has converged
when the mean square deviation $\int{dt \left[Q_n(t)-Q_{n-1}(t)
\right]^2}$ of the distribution $Q_n$ and its predecessor
$Q_{n-1}$ deviate by less than $10^{-4}$. We check that the values
of $z_{\rm low,up}$ do not influence our results and that all the
previous results at the QH transition as in Refs.\ \cite{CaiRSR01}
are reproduced. The full width at half maximum of the FP
distribution $Q_{\rm c}(z)$ is about $5$ \cite{CaiRSR01}.

\section{FP distributions}
\label{sec-FP}

In Figs.\ \ref{fig-P_RL_linlog} and \ref{fig-P_RH}, we show the FP
distributions for $R_{\rm L}$ and $R_{\rm H}$, respectively. As
seen previously \cite{PryA99}, both distributions are manifestly
non-Gaussian with rather long tails. Indeed, $P(R_{\rm L})$ can be
fitted by a log-normal distribution as indicated in Fig.\
\ref{fig-P_RL_linlog} \cite{PryA99}. The FP distribution $P(R_{\rm
H})$ exhibits a pronounced peak at $R_{\rm H}=1$ ($h/e^2$ in SI
units) and a small shoulder around $R_{\rm H}=0$. The small weight
for negative Hall resistances shows that in the coherent RG
structure chosen, the large fluctuations in $P(t)$ can result in
an apparent reversal of the Hall voltage.

\begin{figure}[b]
  \twofigures[width=\twofigwidth]{./fig-P_RL_linlog.eps}{./fig-P_RH-new.eps}
\caption{\label{fig-P_RL_linlog}
  Distribution of the longitudinal resistance $R_{\rm L}$. The $\Box$ symbols indicate the FP distribution, the solid line is a fit
  to a log-normal distribution and the dot-dashed and dashed lines show the
  distribution after $n=15$ RG iterations into the conductance regimes
  $G(z_0 < 0)\rightarrow 0$ and $G(z_0 > 0) \rightarrow 1$. Only every
  5th data point is shown for $P(R_{\rm L})$.}
\caption{\label{fig-P_RH}
  Distribution of the Hall resistance $R_{\rm H}$.  The FP distribution has been
  shaded to zero. For the $G(z_0 < 0)\rightarrow 0$ and $G(z_0 > 0)
  \rightarrow 1$ regimes, we have indicated the distributions after
  $n=16$ RG iterations by bold dashed and dot-dashed lines. }
\end{figure}

\section{The plateau-insulator transition}
\label{sec-PIT}

In order to model the transition into the insulating regime, we
now shift the initial distribution $Q_0(z) \rightarrow Q_0(z-z_0)$
by a small $z_0$. For $z_0 < 0$ and $z_0 > 0$, the RG flow will
then drive the distributions into the insulating, $G \rightarrow
0$, and plateau, $G\rightarrow 1$, regimes, respectively. In
Figs.\ \ref{fig-P_RL_linlog} and \ref{fig-P_RH}, we show the
resulting distributions after many RG steps.
For $P(R_{\rm L})$, we see in Fig.\ \ref{fig-P_RL_linlog} that the
flow towards $G\rightarrow 1$ results in a decrease of large
$R_{\rm L}$ events, whereas conversely, the regime $G \rightarrow
0$ leads to an increase in large $R_{\rm L}$ values and a decrease
of the maximum value in $P(R_{\rm L})$.
For $P(R_{\rm H})$, Fig.\ \ref{fig-P_RH} shows that the plateau
regime $G\rightarrow 1$ gives a highly singular peak at the
dimensionless quantized Hall value $1$, corresponding to a perfect
Hall plateau. On the other hand, the insulating regime $G
\rightarrow 0$ shows an increase of weight in the tails of
$P(R_{\rm H})$ and the eventual obliteration of any central peak.

In order to extract an averaged $R_{\rm L}$ and $R_{\rm H}$ from
these non-standard distributions, we should now select an
appropriate mean $\langle \cdot \rangle$ that characterizes and
captures the essential physics and allows comparison with the
experimental data. The precise operational definition is also
important as it corresponds to different possible experimental
setups. Therefore, we consider several means: (i) arithmetic
$\langle R \rangle_{\rm ari}= \sum_{i} R_i/N$, (ii)
geometric/typical $\langle R \rangle_{\rm typ}= \exp \sum_{i} \ln
R_i/N$, (iii) median (central value) $\langle R \rangle_{\rm
med}$, where $N$ denotes the number of samples ($\gtrsim 10^8$) in
each case. The median and the typical mean (and their variances
\cite{ZulS01}) are less sensitive to extreme values than other
means (such as, e.g.\ root-mean-square and harmonic mean) and this
makes them a better measure for highly skewed and long-tailed
distributions such as $P(R_{\rm L})$ and $P(R_{\rm H})$ in the
insulating regime. In the plateau regimes, the distributions are
less skewed, particularly for $R_{\rm H}$ and the difference in
the means becomes less important.

We are left with determining in which operational order to apply
the averaging procedure. For $R_{\rm L}$, as measured via Ohm's
law (\ref{eq-RL}) as a ratio, it is obvious that the appropriate
average should be ${R}_{\rm L} = \frac{1}{\langle G \rangle} - 1$
(and not $\langle\frac{1}{G}\rangle -1$). For $R_{\rm H}$, the
situation is less straightforward due to the definition of $U_{\rm
H}$ in (\ref{eq-VH}). A simple average is $\langle U_{\rm H}
\rangle$, i.e. using the appropriate $P(R_{\rm H})$. Similar to
the experimental procedure, we can also estimate $U_{\rm H}$ via
$\langle \mu_{15} - \mu_{42} \rangle$ for each $B$ field direction
separately. In Ref.\ \cite{ZulS01}, it has been suggested that a
more appropriate average $\langle U_{\rm H}\rangle^{*}$ can be
constructed from $\langle \mu_{15} \rangle - \langle \mu_{42}
\rangle$. This later procedure corresponds to measuring the
voltage drop between positions $\mu_{15}$ and $\mu_{42}$ in Fig.\
\ref{fig-RGstruct} by separately measuring the individual voltages
with respect to ground and then recombining them.

\begin{figure}[th]
  \twofigures[width=\twofigwidth]{./fig-RHRL-loglog-new_mod.eps}{./fig-RHRL-linlin-new_mod.eps}
\caption{\label{fig-RHRL-loglog}
  Dependence of averaged $R_{\rm H}$ on averaged $R_{\rm L}$ for various means.
The dashed line describes the divergence $R_{\rm H} \propto R_{\rm
L}^{\kappa}$ of the typical mean $\langle R_{\rm
H}\rangle^{*}_{\rm typ}$, the median $\langle R_{\rm
H}\rangle^{*}_{\rm med}$ and the most probable value of $P(R_{\rm
H})$.
  The variance of the arithmetic mean diverges as $R_{\rm L}\rightarrow \infty$ and
  $\langle R_{\rm H}\rangle_{\rm ari}$ is no longer a useful characteristic of
  $P(R_{\rm H})$.
  The horizontal dotted line indicates $R_{\rm H}=1$.
  }
\caption{\label{fig-RHRL-linlin}
  Plot of the same data as in Fig.\ \protect\ref{fig-RHRL-loglog} but on
  a linear scale and for experimentally accessible resistance values.
  Here the almost quantized behavior for the most probable value of $P(R_{\rm H})$
  becomes even more pronounced.
The horizontal dotted line indicates $R_{\rm H}=1$.}
\end{figure}
In Figs.\ \ref{fig-RHRL-loglog} and \ref{fig-RHRL-linlin} we show
the resulting dependence $R_{\rm H}(R_{\rm L})$ when these
averaging definitions are being used. These figures are our main
result.
In the plateau regime $R_{\rm L}<1$, the nearly constant behavior
of all $R_{\rm H}$ averages is as expected. For $R_{\rm L}
\rightarrow \infty$, we get a divergent $R_{\rm H}$ when using the
median and typical means as suggested by Refs.\
\cite{PryA99,ZulS01} to compute both $R_{\rm L}$ and $R_{\rm H}$.
This divergence can be captured by a power-law $R_{\rm H} \propto
R_{\rm L}^{\kappa}$ with $\kappa \approx 0.26$.
The arithmetic mean for large $R_{\rm L} \gg 1$ quickly becomes
instable and no useful information can be inferred.

Reducing the information to the experimentally more relevant
resistance regime of a few times $h/e^2$, we replot the $R_{\rm
L}$ and $R_{\rm H}$ data in Fig.\ \ref{fig-RHRL-linlin}. In
addition to the three means above, we also show the behavior of
the most-probable value $\hat{R}_{\rm H}$ at which $P(R_{\rm H})$
has a maximum. This estimate $\hat{R}_{\rm H}(R_{\rm L})$ appears
relevant in the experimental setup where $10^{8}$ different
samples cannot be easily measured and the full distribution
functions cannot be constructed in similar detail.
Most importantly, for the range of $R_{\rm L}$ values shown in
Fig.\ \ref{fig-RHRL-linlin}, the value of $\hat{R}_{\rm H}$
deviates only slightly from its quantized value $1$ at the
transition and $R_{\rm L}+R_{\rm H}\approx R_{\rm 2t}$
\cite{PelSCS03}. Therefore, the experimental estimate of $R_{\rm
H}$ appears to support the notion of the quantized Hall insulator.
Indeed, the deviations from $1$ are less than $10\%$ until $R_{\rm
L} \sim 25$. However, going back to Fig.\ \ref{fig-RHRL-loglog},
we see that in the strongly insulating regime $R_{\rm L}
\rightarrow \infty$ also $\hat{R}_{\rm H}$ diverges with a
power-law that is well-described by $\langle R \rangle^*_{\rm
typ}$. We emphasize that fluctuations in $\hat{R}_{\rm H}$ for
large $\hat{R}_{\rm L} \gg 10^5$ are due to numerical inaccuracies
in $P(R_{\rm H})$ and decrease upon further increasing the number
of samples.

The results for $R_{\rm L}$ and $R_{\rm H}$ in the localized
regime can be very well described by an exponential scaling
function with finite-size correction \cite{PryA99}
$
  R_{\rm L,H}(2^n,z) \propto 2^{\gamma n} \exp [ {2^n}{\xi^{-1}_{\rm L,H}(z)}]
$.
Plotting $\xi_{\rm L,H}$ as a function of small perturbation
$z_0$, we find $\xi_{\rm L,H}(z_0) \propto z_0^{-\nu_{\rm L,H}}$
with $\nu_{\rm L,H} \approx 2.35$ as shown in Fig.\
\ref{fig-xi_z0_loglog}. Thus we recover the universal divergence
of the localization length $\xi$ even when using the RG in the
insulating regime \cite{Huc92}.
\begin{figure}
 \twofigures[width=\twofigwidth]{./fig-xi_z0_loglog.eps}{./fig-SxxSxy_linlin_new_mod.eps}
\caption{\label{fig-xi_z0_loglog}
  Dependence of the localization lengths $\xi_{L}$ and $\xi_{H}$ on
  the initial shift $z_0$ using the ansatz (8) of \cite{PryA99}. We
  obtain a reasonable estimate of the critical exponent $\nu$.}
\caption{\label{fig-SxxSxy_linlin}
  The semi circle law for different means as in Fig.\ \ref{fig-RHRL-linlin}.
  The most probable values ($\circ$) show perfect semi-circle (solid line) behaviour.
  The thin dotted line denotes $\sigma_{\rm L}={1}/{2}$.}
\end{figure}
An equally reliable estimate of the irrelevant exponent $\gamma$
appears not possible for our data.
As usual, the 4-terminal resistances can be converted into the
respective conductances via $
\sigma_{\rm L,H} = {R_{\rm L,H}}/\left({R_{\rm L}^2 + R_{\rm
H}^2}\right)$.
For these conductances, one expects the semi-circle law
$
    \sigma_{\rm L}^2 +
    \left(\sigma_{\rm H}-{1}/{2}\right)^2 = \left( {1}/{2}
    \right)^2
$ to hold \cite{RuzF95}. In Fig.\ \ref{fig-SxxSxy_linlin}, we show
that the most-probable values capture the overall shape and
symmetry properties best \cite{HilSST98,PelSCS03}, the other
averages show pronounced deviations. We also note that the
relation $R_{\rm L}(z_0)=1/R_{\rm L}(-z_0)$ is obeyed by all means
\cite{ShaHLT98}. This is a consequence of the reflection symmetry
of $P(G)$ \cite{CaiRSR01}.

\section{Conclusion}
\label{sec-concl} We have shown in a quantum coherent calculation
that the insulating Hall behavior of $R_{\rm H}(R_{\rm
L}\rightarrow\infty)$ is dominated by the power-law divergence
$R_{\rm H} \propto R_{\rm L}^{\kappa}$. However, up to previously
experimentally reached $R_{\rm L}\leq 10$
\cite{HilSST98,LanPVP02,PelSCS03}, the deviation from a quantized
$R_{\rm H}=1$ is very small and the onset of the divergence for
$R_{\rm L}\gg 10$ is yet to be explored.

Stimulating discussions with B.\ Huckestein, B.\ Muzykantskii, M.\
E.\ Raikh, and U.\ Z\"{u}licke are gratefully acknowledged. This
research is supported by the DFG within the
Schwer\-punkt\-pro\-gramm ``Quanten-Hall-Systeme'' and the
SFB~393.


\end{document}